# Evaluating the Building Blocks of a Dynamically Adaptive Systematic Trading Strategy

Sonam Srivastava, Mentor – Ritabrata Bhattacharyya

*WorldQuant University*

## Abstract

*Financial markets change their behaviours abruptly. The mean, variance and correlation patterns of stocks can vary dramatically, triggered by fundamental changes in macroeconomic variables, policies or regulations. A trader needs to adapt her trading style to make the best out of the different phases in the stock markets. Similarly, an investor might want to invest in different asset classes in different market regimes for a stable risk adjusted return profile. Here, we explore the use of State Switching Markov Autoregressive models for identifying and predicting different market regimes loosely modelled on the Wyckoff Price Regimes of accumulation, distribution, advance and decline. We explore the behaviour of various asset classes and market sectors in the identified regimes. We look at the trading strategies like trend following, range trading, retracement trading and breakout trading in the given market regimes and tailor them for the specific regimes. We tie together the best trading strategy and asset allocation for the identified market regimes to come up with a robust dynamically adaptive trading system to outperform simple traditional alphas.*

# Table of Contents

# 1. Introduction

As macroeconomic conditions shift from steady low-volatility economic growth regimes to panic-driven high volatility regimes, the markets transition into advancing and declining trends which in common terminology is categorized as bull and bear markets. The regimes are characterized by varying mean, variance and correlation patterns of asset returns and are not clearly defined. Simple systematic trading models based on linear assumptions can give very good results in a specific market regime but as the behaviour pattern shifts the simple linear models crumble. To tackle this non-linearity and cyclicity, regime switching models have gained a lot of popularity in Financial modelling in recent times. One of the most popular models for this are the Markov Switching Autoregressive models. In the seminal work on *Regime Switching Models* (Hamilton, 2005), Hamilton successfully introduced these models to forecast economic recessions and following him numerous researchers have demonstrated the use of Markov Switching models for regime shift detection in financial time series modelling.

Using the philosophy of a Markov Switching Auto Regressive model to predict market regime we adopt the Markov switching heteroscedasticity framework to decompose stock return into permanent and transitory components and predict the probability of two variance states in stock returns similar to the model presented in *Mean Reversion in Stock Prices?* (Myung Jig Kim, 1991). This model gives us the smoothed probability of low and high variance states which when clubbed with the long-term triangular moving average regimes provides us strong segmentation of market regimes against which we can study the performance of various asset classes and stock market sectors and various trading techniques or alphas.

Adapting to different regimes adds a new dimension to the performance of investments and trading strategies. For example, in a bull market trend following and momentum strategies on the stock market are enormously attractive, while in the times of doom one wants to switch to shorting the market or moving to asset classes like fixed income and gold to outperform the market.

Our results find validation in the wisdom imparted by professional traders and portfolio managers who prefer stocks in bull markets and government securities or gold in bear markets and buy retracement in an advancing trend and sell resistances in a declining cycle while sticking to range bound trading in consolidation phases. We intend to engrain this wisdom in a systematic strategy using the Markov Switching models to build a robust dynamically adaptive trading system.

## I. Markov Switching Auto Regressive Model

Modelling asset prices as an autoregressive time series, the current state of the series $y_t$ depends on the autoregressive component $y_{t-1}$, random shocks $\epsilon_t$ and a regime process $S_t$.

Regimes are modelled through a discrete variable, $S_t \in \{0,1,...,k\}$, tracking the particular regime inhabited by the process at a given point in time.

Although regimes could affect the entire distribution, regimes are often limited to affect the intercept $\mu_{S_t}$, autocorrelation, $\phi_{S_t}$, and volatility, $\sigma_{S_t}$, of the process:

*Equation 1 Autoregressive Regime Switching*

$$y_t = \mu_{S_t} + \phi_{S_t} y_{t-1} + \sigma_{S_t} \epsilon_t$$

Where $\epsilon_t \sim iid(0,1)$

The governing dynamics of the underlying regime, $S_t$, are assumed to follow a homogenous first-order Markov chain,

*Equation 2 Regime Switching Probabilities*

$$\Pi_{i,j} = \Pr(S_t = i | S_{t-1} = j) = p_{ij}$$

Estimation of the transition probabilities $\Pi_{i,j}$ is usually done by a **Maximum Likelihood Estimator**.

This further leads to calculation of smoothed probabilities and expected durations. (Hamilton J. D., 2014)

## 2. Regime Shift Model

Richard Wyckoff proposed a four-stage Market cycle based on the behaviour of smart investors in markets governed by supply and demand. Wyckoff trading cycles can provide a very good explanatory framework of market regimes and price movements (Wyckoff, 1931) The idea is that prices move between the phases of strong uptrend (advance), strong downtrend (decline) and range bound consolidation periods (accumulation and distribution) based on supply-demand dynamics and the movement of smart money.

We use a Two-Variance Markov Switching Model (Myung Jig Kim, 1991) which demonstrates estimation with regime heteroskedasticity (switching of variances) and no mean effect to classify the market into low variance and high variance regimes. We then superimpose a long term moving average regime to further classify the regimes into bearish and bullish regimes.

A triangular moving average (Zakamulin, 2015) which performs double smoothing of the stock prices (with 250 days period) is used. The triangular moving average is chosen over simple moving averages and exponential moving averages as the triangular moving average is double smoothed and thus would inhibit small horizon whipsaws.

A Keltner Channel (Keltner, 2010) based on a 20 days Average True Range (ATR) of stock prices on top of the moving average to only trigger a regime shift when the price moved away from the moving average by one ATR to avoid unnecessary whipsaws again.

## 3. Modelled Regimes

Modelling two variance regimes in the returns of the **S& 500 Index** (from 2000 to 2017) using the Markov Switching Autoregressive Model we obtain the following smoothed probabilities of high variance and low variance periods (*Figure 1*).

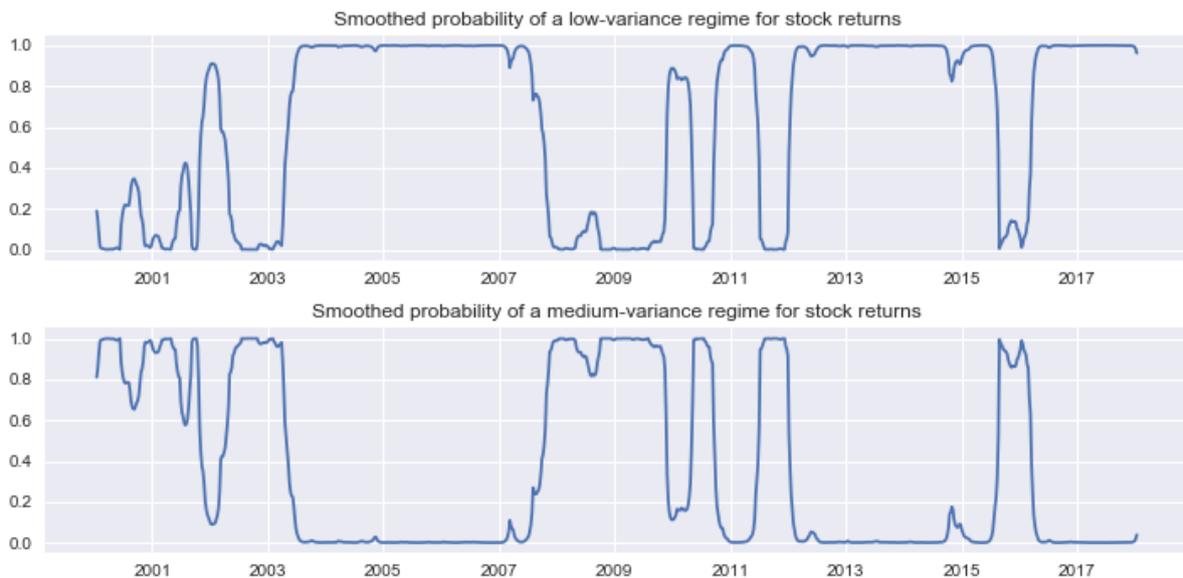

*Figure 1 Variance Regime Probabilities*

|  | Regime 0 parameters | | | | | |
| --- | --- | --- | --- | --- | --- | --- |
|  | coef | std err | z | P>\|z\| | [0.025 | 0.975] |
| sigma2 | 0.0002 | 1.16e-05 | 13.889 | 0.000 | 0.000 | 0.000 |
|  | Regime 1 parameters | | | | | |
|  | coef | std err | z | P>\|z\| | [0.025 | 0.975] |
| sigma2 | 0.0009 | 8.68e-05 | 9.921 | 0.000 | 0.001 | 0.001 |
|  | Regime transition parameters | | | | | |
|  | coef | std err | z | P>\|z\| | [0.025 | 0.975] |
| p[0->0] | 0.9867 | 0.006 | 165.930 | 0.000 | 0.975 | 0.998 |
| p[1->0] | 0.0236 | 0.012 | 1.930 | 0.054 | -0.000 | 0.047 |

*Table 1 Regime Transfer Parameters*

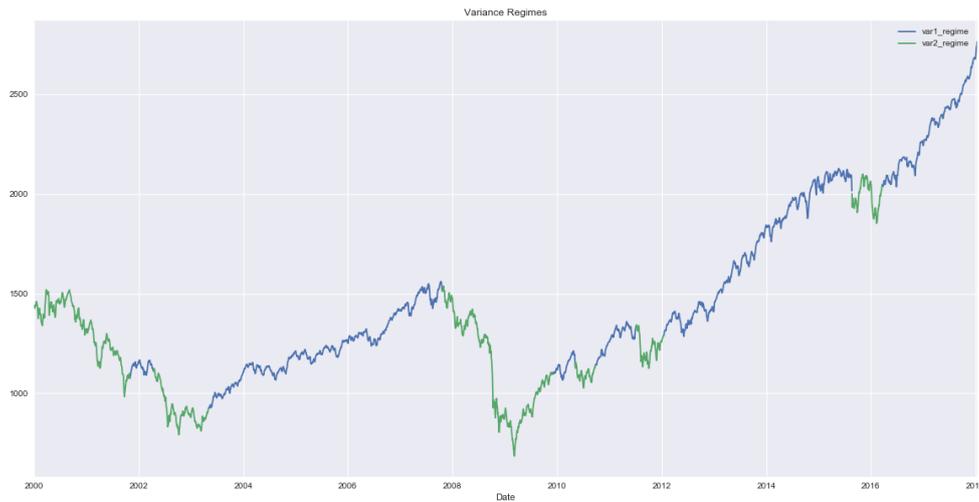

*Figure 2 Variance Regimes Visualized*

250-day Triangular Moving Averages are used to further divide the regimes to bullish or bearish classifications (*Figure 3*). Using the combination of variance and moving average regimes, few small duration regimes were observed – which would cause whipsaws, thus Keltner channels based on 20 days Average True Range are further applied to confirm a regime shift.

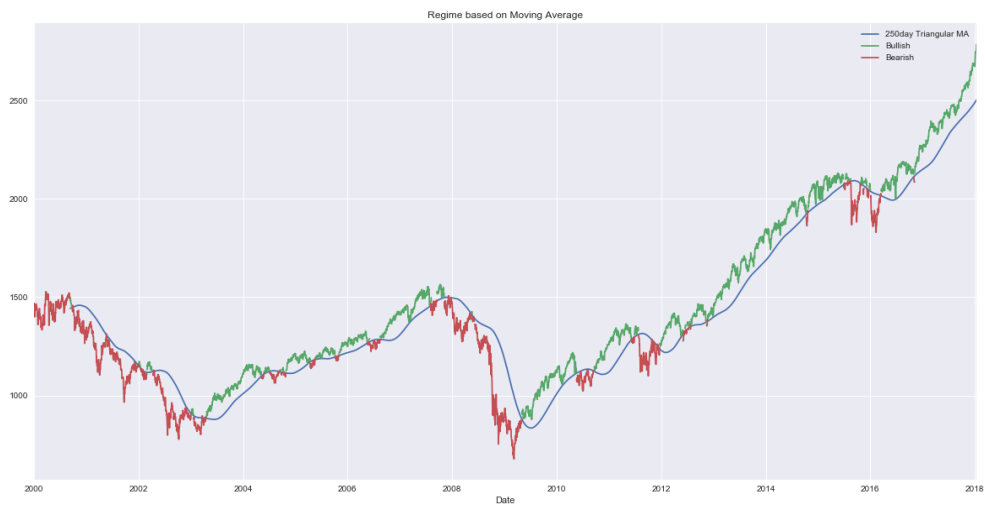

*Figure 3 Moving Average Regimes Visualized*

We use a combination of the variance regimes and moving average regimes to get 4 distinct market regimes to study the strategy returns on (*Figure 4*).

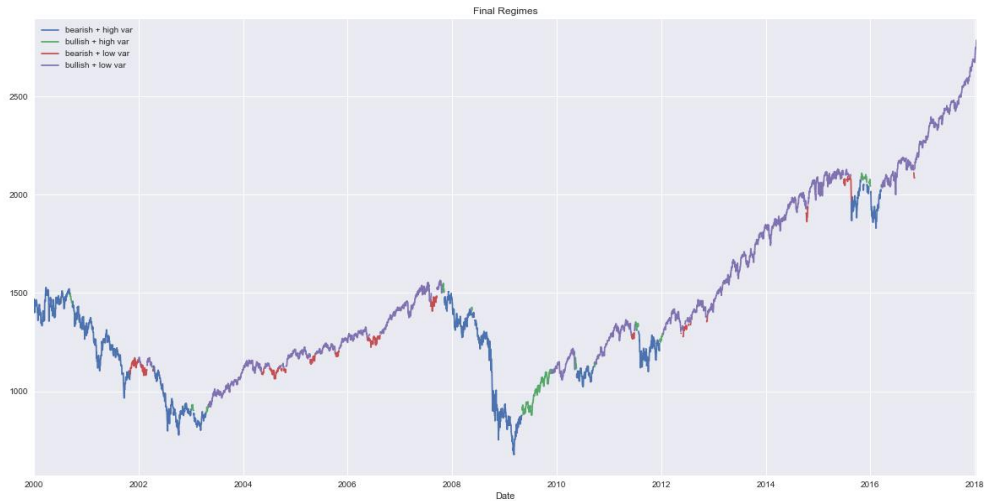

*Figure 4 Final Regimes Visualized*

The descriptive stats for the regime lengths are:

*Table 2 Regime Lengths*

| count | 81 |
|---|---|
| mean | 77 |
| std | 128 |
| min | 2 |
| 25% | 9 |
| 50% | 15 |
| 75% | 78 |
| max | 687 |

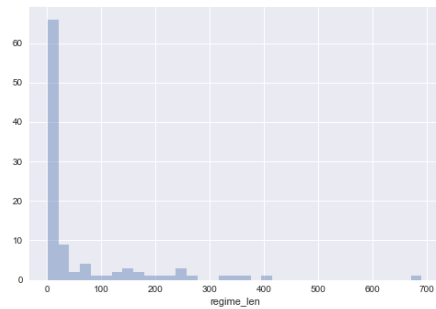

*Figure 5 Regime Lengths*

The returns in the segregated regimes are described as:

|  | *Mean Daily Return* | *Daily Standard Deviation* | *Number of Days* | *Mean Regime Length* |
|---|---|---|---|---|
| bearish + high variance | -0.041% | 1.533% | 2011 | 145 |
| bearish + low variance | 0.011% | 0.821% | 564 | 24 |
| bullish + high variance | 0.085% | 0.966% | 363 | 25 |
| bullish + low variance | 0.039% | 0.578% | 3649 | 120 |

*Table 3 Return Characteristics in various regimes*

To get a better picture of the price behaviour in the classified regimes, we plot the largest classified time periods in the four regimes (*Figure 6*).

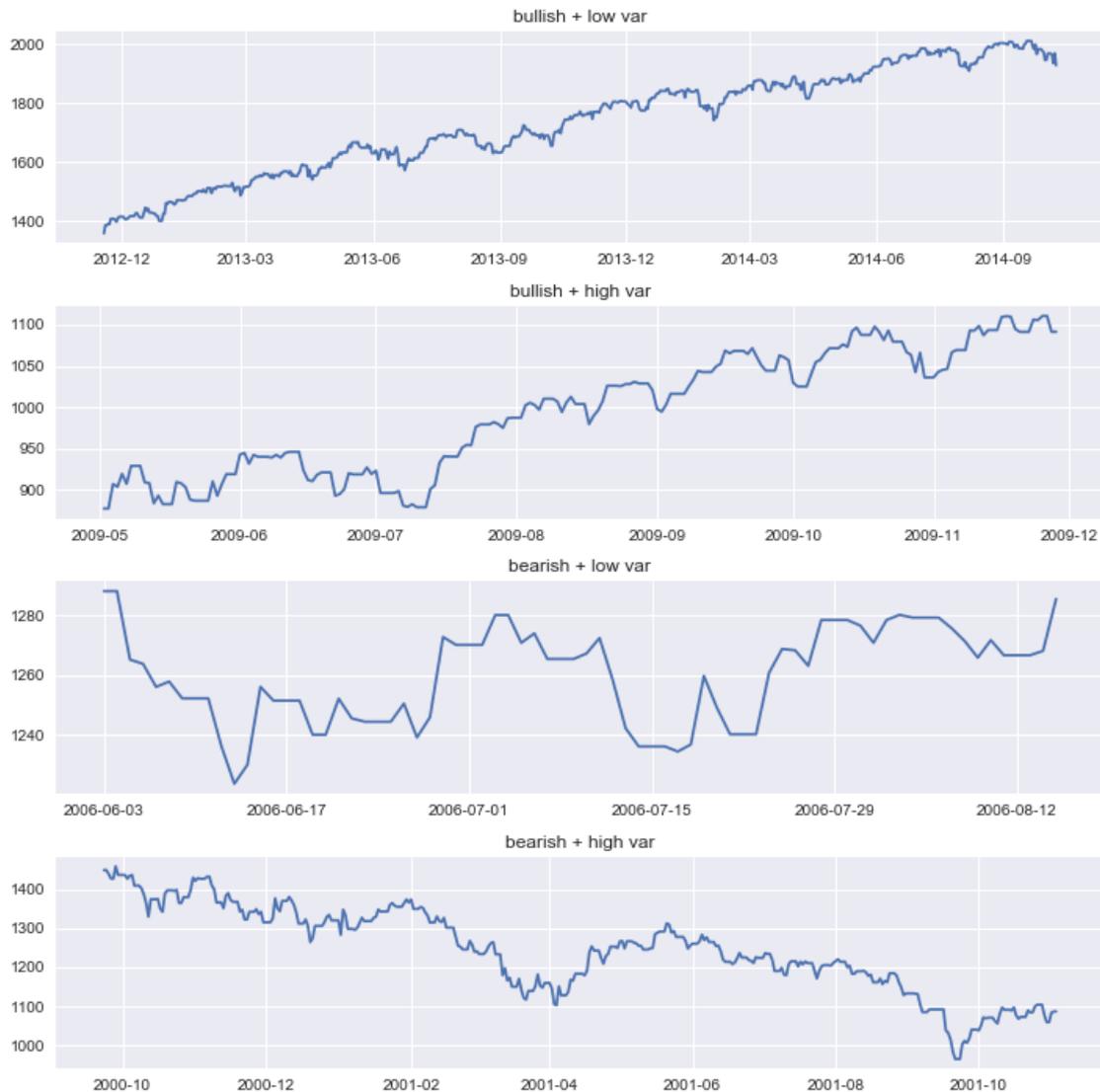

*Figure 6 Representative price patterns in different regimes*

It is very visible from the figure that the four classified regimes have very distinct price behaviours.

- *Bearish + high variance* regime has steadily declining prices.
- *Bullish + low variance* regime has steadily increasing prices.
- *Bearish + low variance* regime has prices which are range bound.
- *Bullish + high variance* regime has increasing prices with higher variance.

It is quite intuitive to expect that different asset classes would have distinct patterns of returns in the given regimes and distinct trading styles would fit in to maximize trading returns in these distinct regimes.

# 4. Asset Class Behaviour[1]

Asset pricing models like CAPM attribute market returns to be a big factor in determining asset prices. It is interesting to look at how asset prices react in different market conditions. (Don U.A. Galagedera, 2005)

The asset classes whose returns we look at are -
- S&P 500 Index
- MSCI World Index
- MSCI Emerging Market Index
- Long Term US Government Securities
- Short Term US Government Securities
- Gold
- Crude Oil

The mean returns of these assets under various market regimes are –

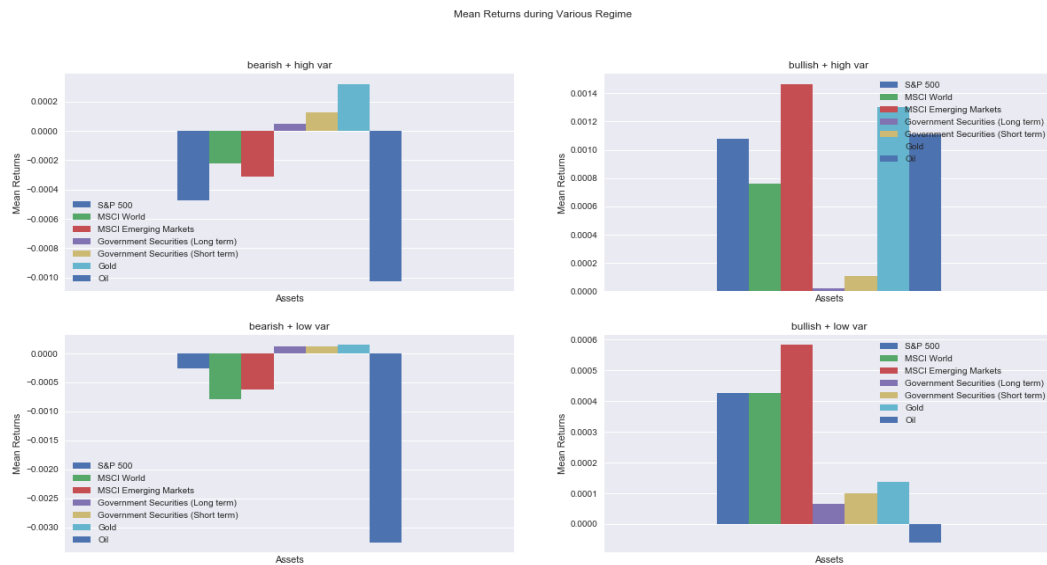

Figure 7 Asset Returns in Different Regimes

|  | bearish + high var | bearish + low var | bullish + high var | bullish + low var |
|---|---|---|---|---|
| **S&P 500** | -0.05% | -0.03% | 0.11% | 0.04% |
| **MSCI World** | -0.02% | -0.08% | 0.08% | 0.04% |
| **MSCI Emerging Markets** | -0.03% | -0.06% | 0.15% | 0.06% |
| **Government Securities (Long term)** | 0.00% | 0.01% | 0.00% | 0.01% |
| **Government Securities (Short term)** | 0.01% | 0.01% | 0.01% | 0.01% |
| **Gold** | 0.03% | 0.02% | 0.13% | 0.01% |
| **Oil** | -0.10% | -0.33% | 0.11% | -0.01% |

Table 4 Asset Returns in Different Regimes

---

[1] *The data sources used for getting the asset prices are in the appendix.*

We can see here notably that the equity indices have positive returns during bullish regimes and fixed income and gold have positive returns during both regimes. Oil has an inconsistent behaviour.

Looking at the correlation pattern of returns in the different regimes, we see that the correlations between asset classes are higher in the high variance regimes.

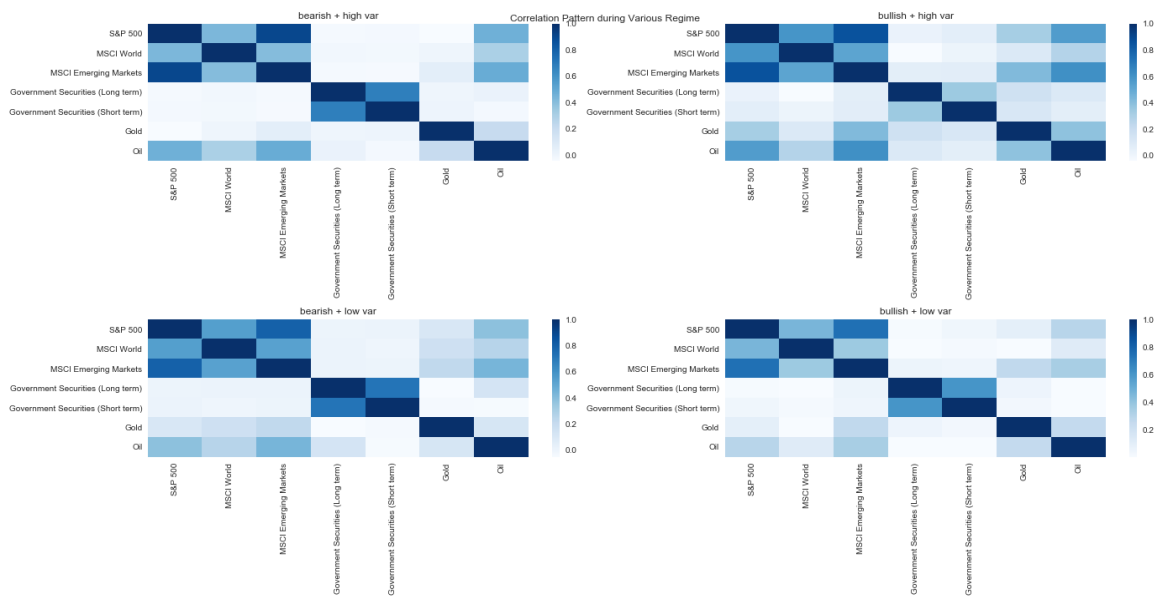

*Figure 8 Correlation between asset returns*

Market Beta also varies with varying market regimes (Roland Shami, 2009) and the effect of this phenomenon on the prices of various market sector indices can be observed (*Figure 9*).

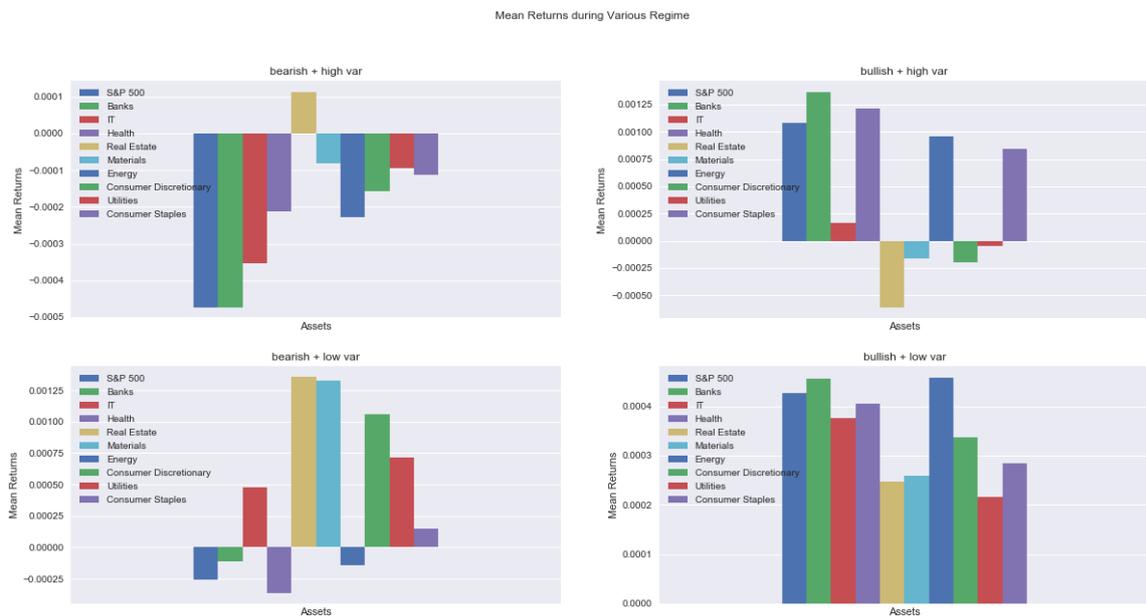

*Figure 9 Sector Returns in Different Regimes*

|         | *bearish + high var* | *bearish + low var* | *bullish + high var* | *bullish + low var* |
|--------:|---------------------:|--------------------:|---------------------:|--------------------:|
| **S&P 500** | -0.05%           | -0.03%              | 0.11%                | 0.04%               |
| **Banks**   | -0.05%           | -0.01%              | 0.14%                | 0.05%               |
| **IT**      | -0.04%           | 0.05%               | 0.02%                | 0.04%               |
| **Health**  | -0.02%           | -0.04%              | 0.12%                | 0.04%               |
| Real Estate | 0.01% | 0.14% | -0.06% | 0.02% |
| Materials | -0.01% | 0.13% | -0.02% | 0.03% |
| Energy | -0.02% | -0.01% | 0.10% | 0.05% |
| Consumer Discretionary | -0.02% | 0.11% | -0.02% | 0.03% |
| Utilities | -0.01% | 0.07% | 0.00% | 0.02% |
| Consumer Staples | -0.01% | 0.01% | 0.08% | 0.03% |

*Table 5 Sector Returns in various regimes*

We can see that the low beta sectors like Energy, Utilities, Consumer Staples and Consumer Discretionary sectors outperform the index during bearish regimes while high beta sectors like Bank and IT outperform during bullish regimes. (Roland Shami, 2009)

Looking at the correlation patterns again we see that the correlations are higher in the high variance regimes than the low variance regimes.

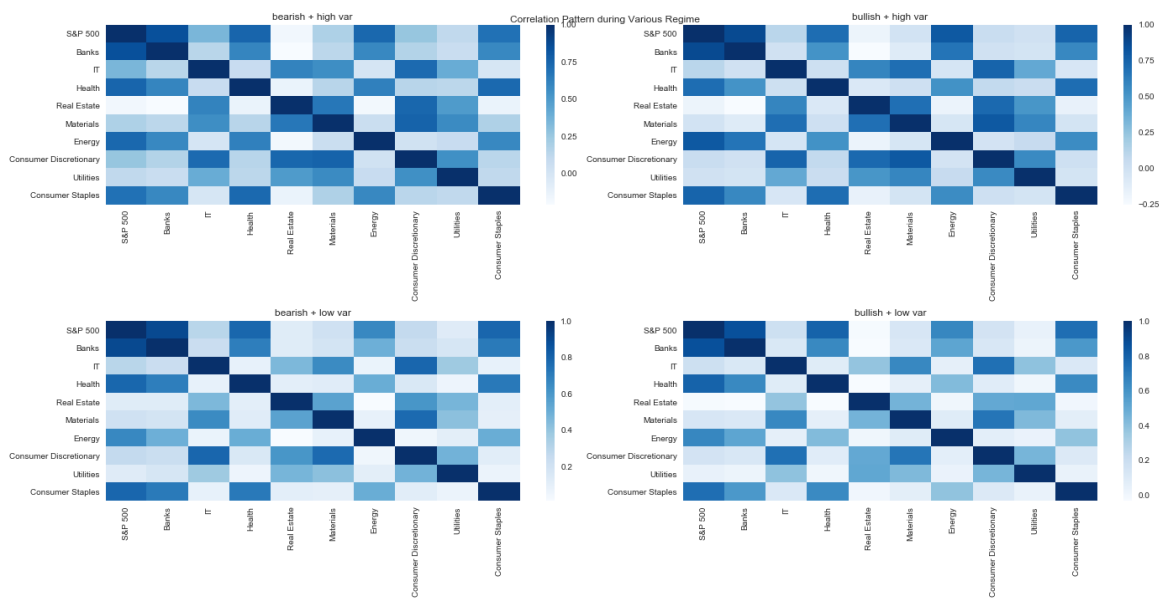

*Figure 10 Correlation pattern in sector returns*

## 5. Trading Strategies - Alphas

Our market regimes loosely fit into the Wyckoff framework as –

- *Bullish Low Variance → Advance*
- *Bearish High Variance → Decline*
- *Bullish High Variance → Accumulation*
- *Bearish Low Variance → Distribution*

As the behaviour of the price trend and variance is very typical in each of these regimes, we can fit very specific strategies to get good risk adjusted returns in each of the regimes. For which purpose we study a set of technical signals and their behaviour in different regimes.

### I. Trading Signal Analysis

We look at the following technical signals for our strategies –

- Bollinger bands
  *Volatility bands placed above and below a moving average*
- Relative Strength Index

*A momentum oscillator that measures the speed and change of price movements*
- Fibonacci Retracements
  *Dividing distance between extreme points in key Fibonacci ratios of 23.6%, 38.2%, 50%, 61.8% and 100%*
- Exponential Moving Averages

The price trend after the chosen technical signal occurs is studied here till 30 days or regime change. Based on these trends, trading strategies are suggested.

### A. Bollinger Bands

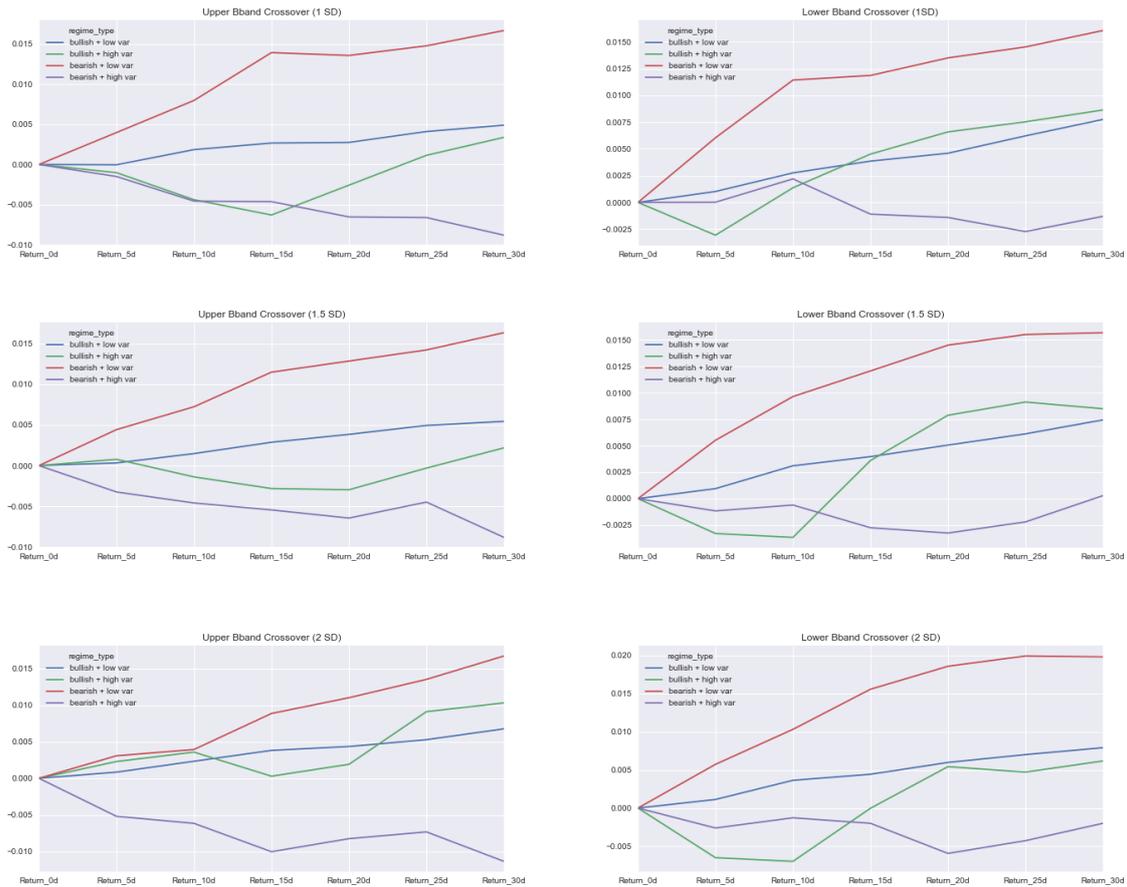

*Figure 11 Bollinger Bands Analysis*

- Upper Bollinger Band Crossover → go long in bearish low variance and bullish low variance regimes.
- Upper Bollinger Band Crossover → go long in bullish high variance regime.
- Upper Bollinger Band Crossover → go short for bearish high variance regime.
- Lower Bollinger Band Crossover → go long in bearish low variance and bullish low variance regimes.
- Lower Bollinger Band Crossover → go short for bearish high variance.

### B. Fibonacci Crossover

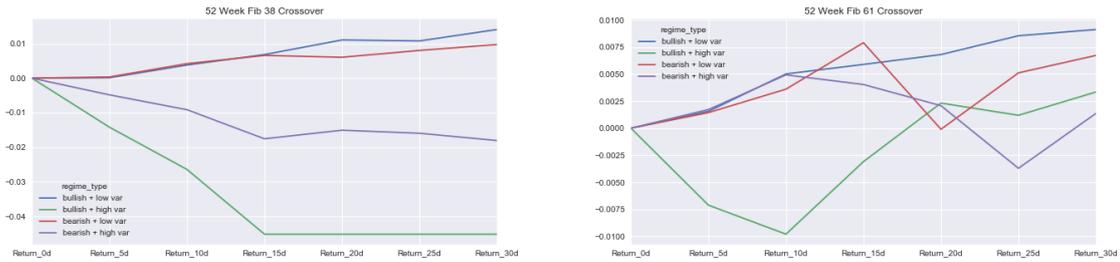

*Figure 12 Fibonacci Signal Analysis*

- Fibonacci 38 level crossover → go long in bullish and bearish low variance regimes.
- Fibonacci 38 level crossover → go short in bullish and bearish high variance regimes.
- Fibonacci 61 level crossover → go long in bullish and bearish low variance regimes.

### C. MACD Crossover

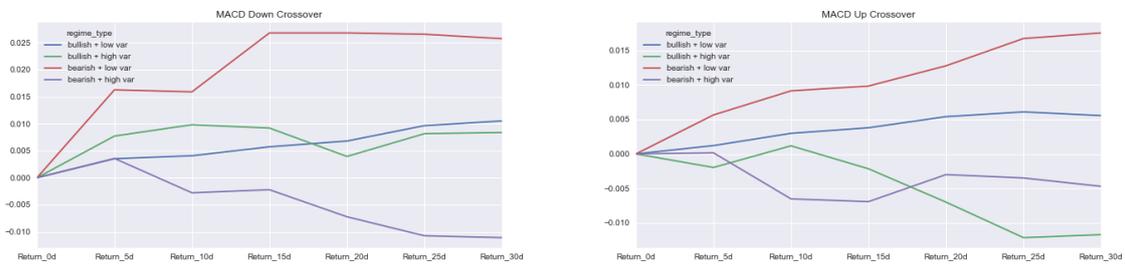

*Figure 13 MACD Signal Analysis*

- MACD low crossover → go long in bearish and bullish low variance regimes.
- MACD low crossover → go short in bearish high variance regimes.
- MACD up crossover → go long in bearish and bullish low variance regimes..
- MACD up crossover → go short in bearish and bullish high variance regimes.

### D. RSI Crossover

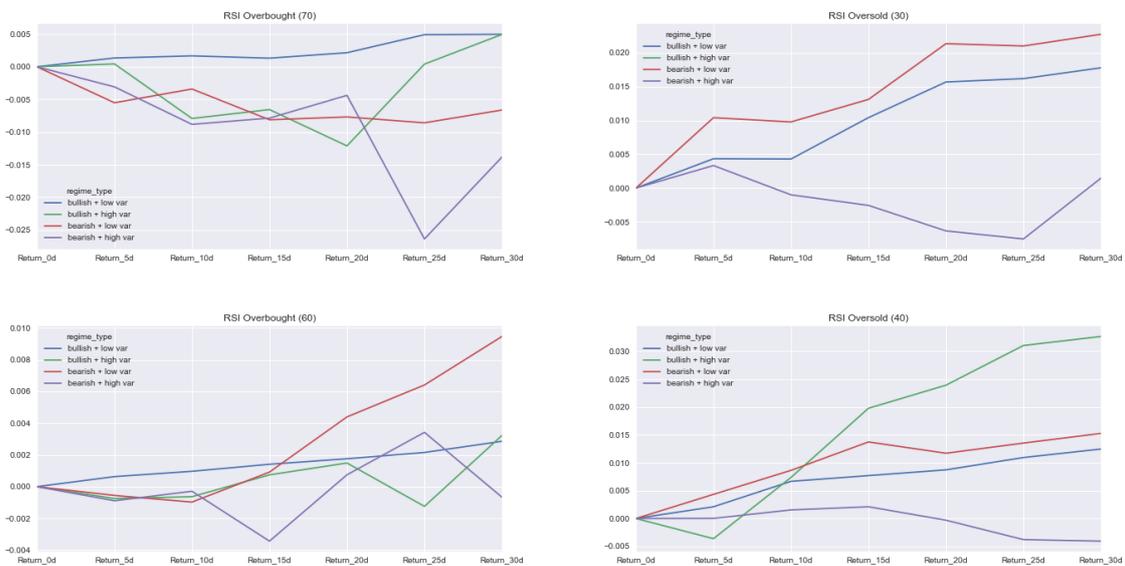

*Figure 14 RSI Signal Analysis*

- RSI 60 crossover → go long in bearish and bullish low variance regimes
- RSI 30 crossover → go long in bearish and bullish low variance regimes.
- RSI 30 crossover → go short in bearish high variance regime.
- RSI 70 crossover → go short in bearish and bullish high and bearish low variance regimes.
- RSI 30 crossover → go long in bearish and bullish low and bullish high variance regimes.
- RSI 30 crossover → go short in bearish high variance regimes.

## II. Tailored Trading Strategies

Using these signals, the trading strategy for each regime can be defined. The exact signals from the above analysis were not used. Some trial was done to get the best fit in each regime based on above analysis.

### A. ADVANCE

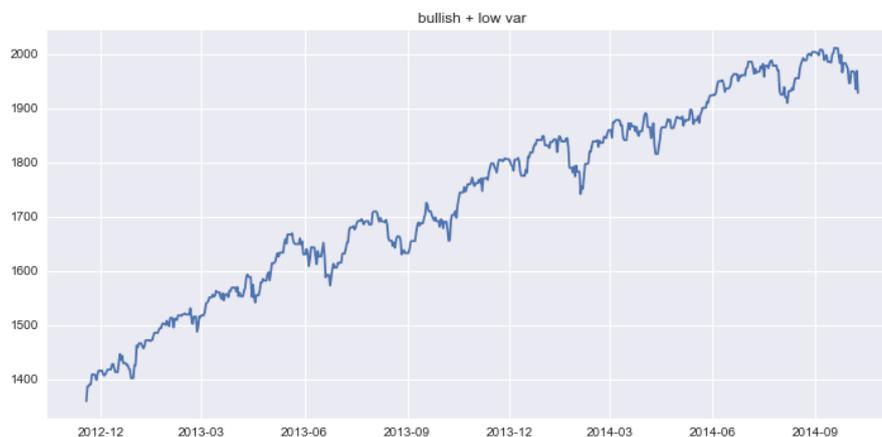

*Figure 15 Representative Price Movement in Advance Regime*

As evidenced by the representative chart for the regime the strategy for trading here is -

- ✓ *Follow Up trend*
- ✓ *Go long on pullbacks and breakouts*

Consequently, the trading signals are:

*Long Signals*

- Price crosses below lower Bollinger band (1.5 SD)
- Relative Strength Index crosses below lower threshold (40)
- Price crosses below 38.2% Fibonacci level

*Exit Long Signals*

- Price crosses above upper Bollinger band (1.5 SD)
- Relative Strength Index crosses below lower threshold (70)
- Price crosses below short-term Exponential Moving Average (10 days)
- Price crosses below 61.8% Fibonacci level
- Take profit at EMA + 3*ATR
- Take loss at EMA – 3*ATR

### B.  ACCUMULATION

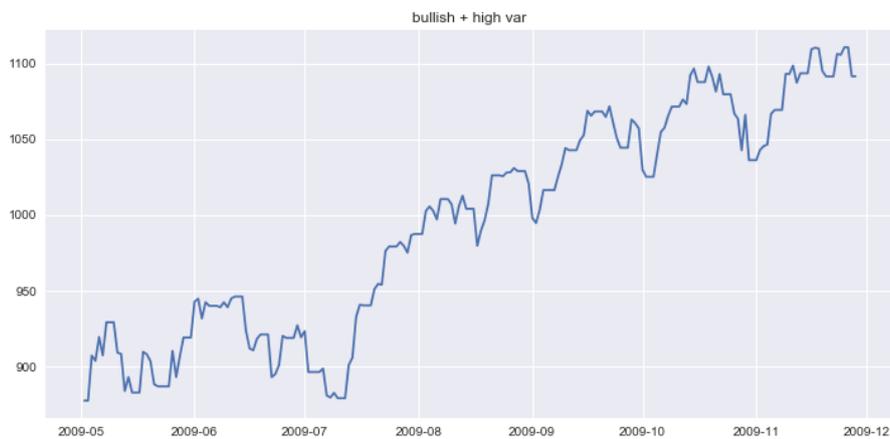

Figure 16 Representative Price Movement in Accumulation Regime

- ✓ Long lows of the range
- ✓ Stop loss above support, Take profit Swing High

*Long Signals*

- Price crosses below lower Bollinger band (1 SD)
- Relative Strength Index crosses below lower threshold (40)
- Price crosses below 38.2% Fibonacci level

*Exit Long Signals*

- Price crosses above upper Bollinger band (1.5 SD)
- Relative Strength Index crosses below lower threshold (70)
- Price crosses below short-term Exponential Moving Average (10 days)
- Price crosses below 61.8% Fibonacci level
- Take profit at EMA + 3*ATR
- Take loss at EMA – 2*ATR

### C.  DECLINE

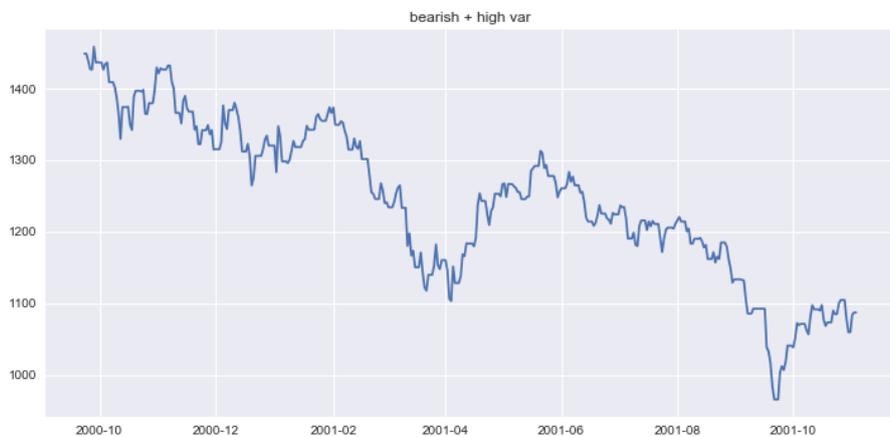

Figure 17 Representative Price Movement in Decline Regime

- ✓ Follow Down trend
- ✓ Go short on pullbacks and breakouts

*Short Signals*

- Price crosses above upper Bollinger band (1.5 SD)
- Relative Strength Index crosses above upper threshold (60)
- Price crosses above short-term Exponential Moving Average (10 days)
- Price crosses above 61.8% Fibonacci level

*Exit Short Signals*

- Price crosses below lower Bollinger band (1 SD)
- Relative Strength Index crosses below lower threshold (20)
- Price crosses below short-term Exponential Moving Average (10 days)
- Take profit at EMA - 5*ATR
- Take loss at EMA + 5*ATR

### D. DISTRIBUTION

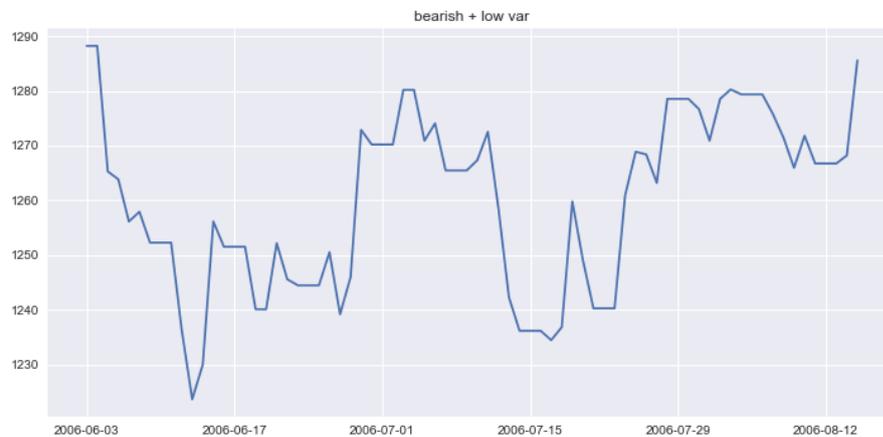

*Figure 18 Representative Price Movement in Distribution Regime*

- ✓ Short highs of the range
- ✓ Stop loss above resistance, Take profit Swing Low

*Short Signals*

- Price crosses above upper Bollinger band (1.5 SD)
- Relative Strength Index crosses above upper threshold (60)
- Price crosses above 61.8% Fibonacci level

*Exit Short Signals*

- Price crosses above upper Bollinger band (1 SD)
- Relative Strength Index crosses below lower threshold (20)
- Take profit at EMA - 3*ATR
- Take loss at EMA + 3*ATR

## 6. Back testing Results[2]

---

[2] *The description of the risk adjusted stats is in the appendix*

Back tests are performed based on the above four alphas tailored for the four regimes and results are analysed in each regime.

*Back tests done to invest initial capital of 1 million dollar and keep reinvesting the profits. A transaction cost of 1 bps is assumed for each trade. Slippage assumed is 1 bps and 1 bps are set as commission. Gap-ups are not considered, and only close prices are used as there is only close prices data available for a lot of the assets used.*

The results of the above alphas applied to the S&P 500 index in the identified regimes are –

### I. ADVANCE

| CAGR % | 6.57 |
|---|---|
| Annualized Risk % | 22.33 |
| Sharpe % | 29.41 |
| Win % | 29.38 |
| Num Trades | 2481 |
| Win to Loss Ratio % | 41.61 |
| Average Daily PL $ | 456.1 |
| Daily Return bps (total) | 6.84 |
| Max Consecutive Losers | 30 |
| Max Drawdown % | -37.82 |
| Lake Ratio | 6.4 |
| Gain to Pain Ratio | 30.4 |

*Table 6 Back test Result - Advance Regime*

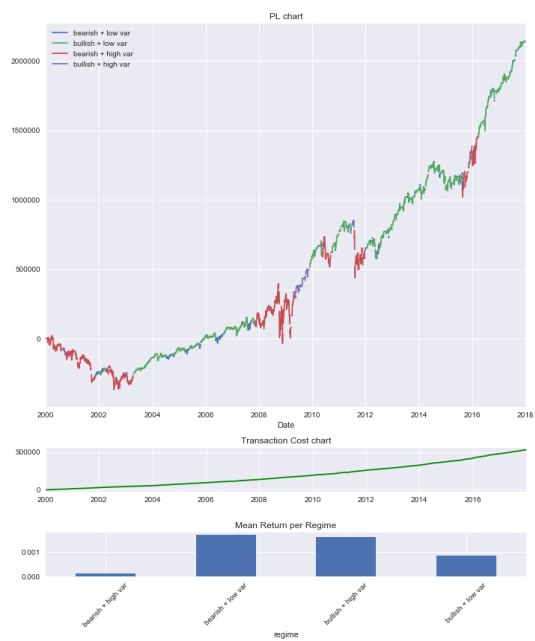

*Figure 19 Back test Result - Advance Regime*

### II. ACCUMULATION

| CAGR % | 6.29 |
|---|---|
| Annualized Risk % | 21.91 |
| Sharpe % | 28.72 |
| Win % | 32.26 |
| Num Trades | 2619 |
| Win to Loss Ratio % | 47.63 |
| Average Daily PL $ | 426.0 |
| Daily Return bps (total) | 6.46 |
| Max Consecutive Losers | 24 |
| Max Drawdown % | -43.22 |
| Lake Ratio | 8.6 |
| Gain to Pain Ratio | 49.0 |

*Table 7 Back test Result - Accumulation Regime*

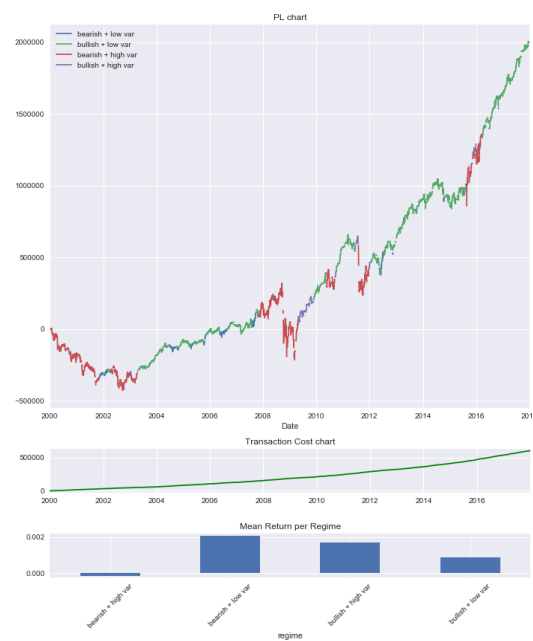

*Figure 20 Back test Result - Accumulation Regime*

## III. DECLINE

| | |
|---|---|
| CAGR % | 0.87 |
| Annualized Risk % | 15.95 |
| Sharpe % | 5.46 |
| Win % | 19.62 |
| Num Trades | 3288 |
| Win to Loss Ratio % | 24.40 |
| Average Daily PL $ | 36.0 |
| Daily Return bps (total) | 2.01 |
| Max Consecutive Losers | 38 |
| Max Drawdown % | -55.21 |
| Lake Ratio | 32.9 |
| Gain to Pain Ratio | 15.8 |

*Table 8 Back test Result - Decline Regime*

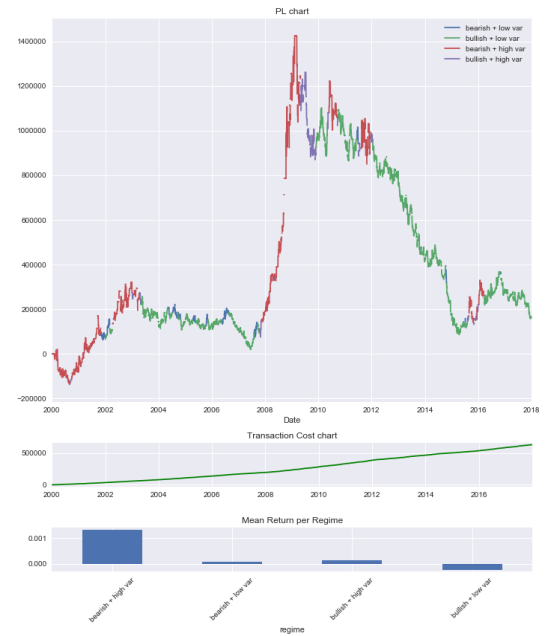

*Figure 21 Back test Result - Decline Regime*

## IV. DISTRIBUTION

| | |
|---|---|
| CAGR % | 1.93 |
| Annualized Risk % | 15.50 |
| Sharpe % | 12.43 |
| Win % | 24.72 |
| Num Trades | 3171 |
| Win to Loss Ratio % | 32.84 |
| Average Daily PL $ | 87.3 |
| Daily Return bps (total) | 2.66 |
| Max Consecutive Losers | 44 |
| Max Drawdown % | -57.72 |
| Lake Ratio | 31.2 |
| Gain to Pain Ratio | 25.5 |

*Table 9 Back test Result - Distribution Regime*

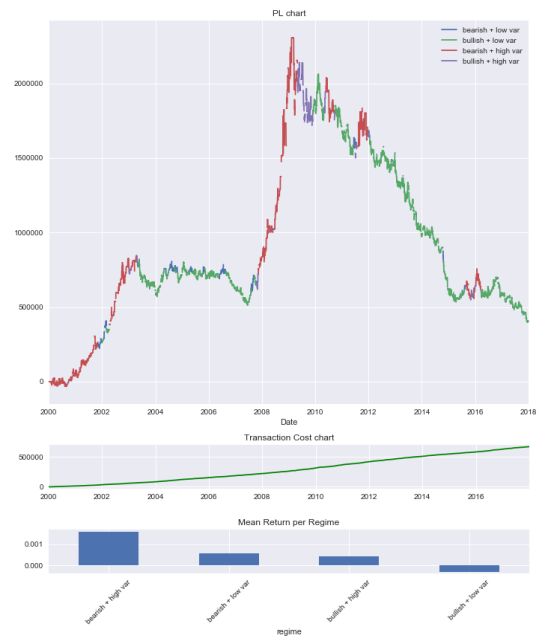

*Figure 22 Back test Result - Distribution Regime*

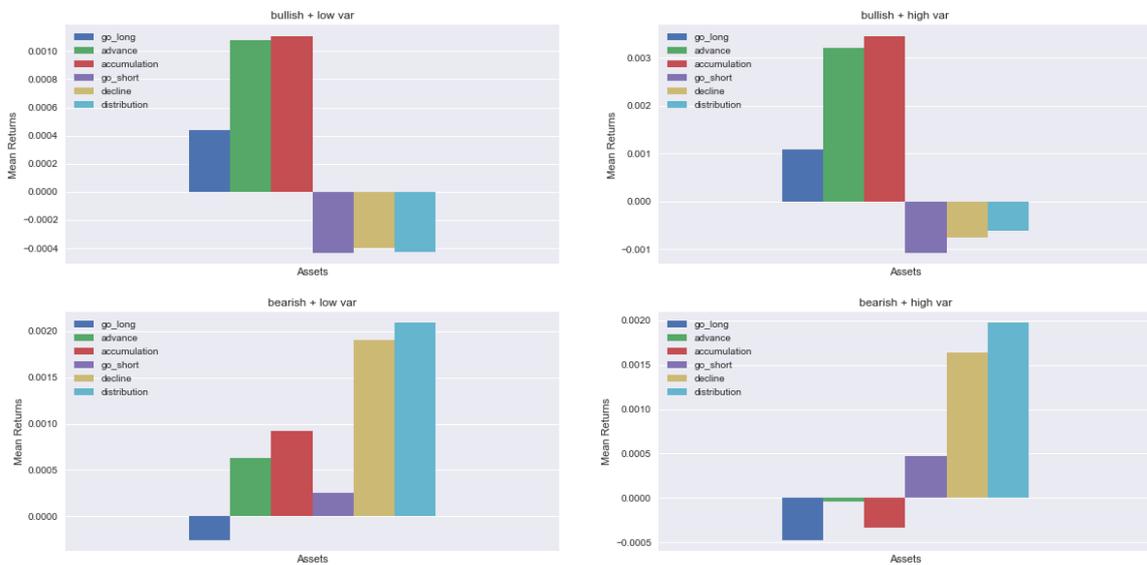

*Figure 23 Performance of Alphas in Different Regimes*

We can see that the alphas perform well in the regimes they are tailored for. The decline and distribution alphas outperform during bearish regimes while the advance and accumulation alphas outperform in bullish regimes. The accumulation alpha is more suited for the high variance regime than the advance alpha.

The bullish alphas have higher returns and Sharpe ratios as most of days in the given time were bullish.

## 7. Dynamically Adaptive Strategy

We combine the four alphas such that the strategy switches to the participation methodology of the given regime based on regime type.

To avoid whipsaws, when switching to a bullish regime the trades are switched only when the market moves 1% in the bullish side and in the bearish regime when the market moved 5% in the bearish side.

The results for the back test are

| CAGR % | 12.54 |
|---|---|
| Annualized Risk % | 18.28 |
| Sharpe % | 68.61 |
| Win % | 30.11 |
| Num Trades | 2836 |
| Win to Loss Ratio % | 43.09 |
| Average Daily PL $ | 1574.3 |
| Daily Return bps (total) | 9.73 |
| Max Consecutive Losers | 124 |
| Max Drawdown % | -21.77 |
| Lake Ratio | 3.0 |
| Gain to Pain Ratio | 22.4 |

*Table 10 Back test results adaptive backtest*

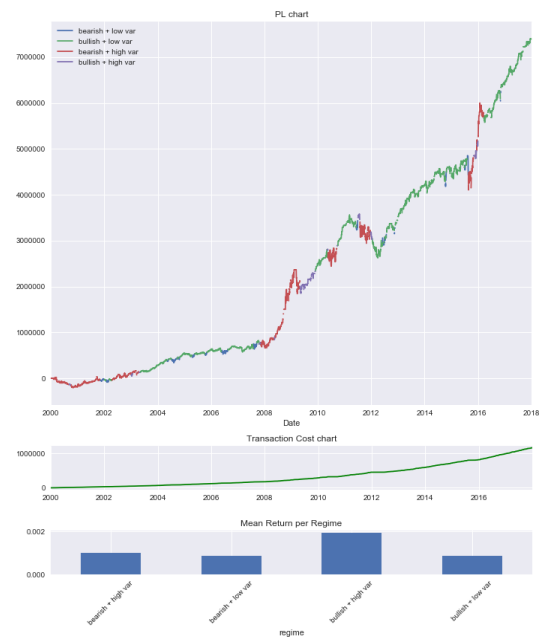

*Figure 24 Back test results adaptive backtest*

Looking at the comparative statistics of the back tests

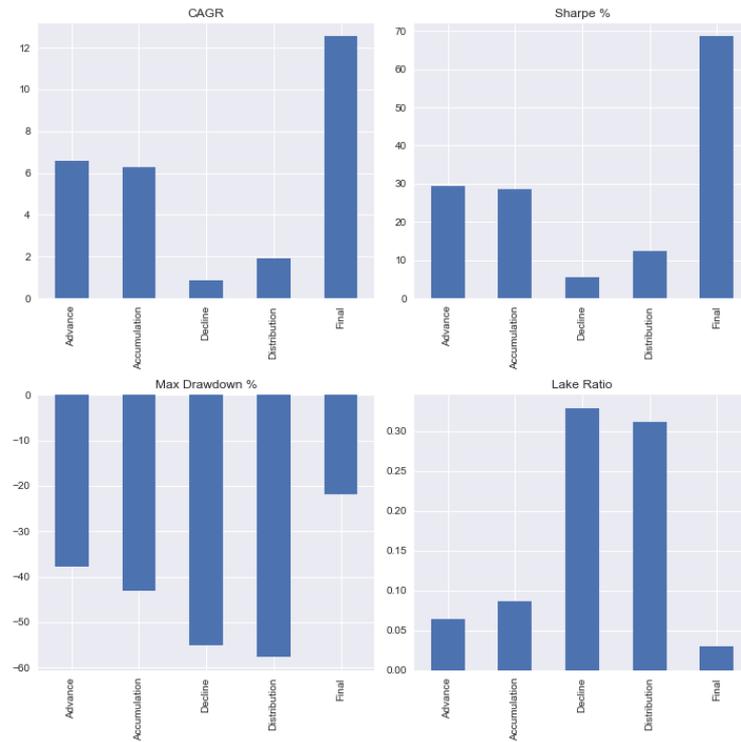

*Figure 25 Performance Comparison*

We can see that the dynamically adaptive back test has outperformed the other back tests drastically in terms of returns with the final strategy giving a CAGR of 12.5% with Sharpe of 68.6% while the best single strategy gave CAGR of 6.57% with Sharpe of 29%. Not only are the absolute returns and risk adjusted return stats are better for the final back test, the critical risk parameters like maximum drawdown and lake ratio are lower compared to the other back tests. The strategy also gives competitive returns in all regimes.

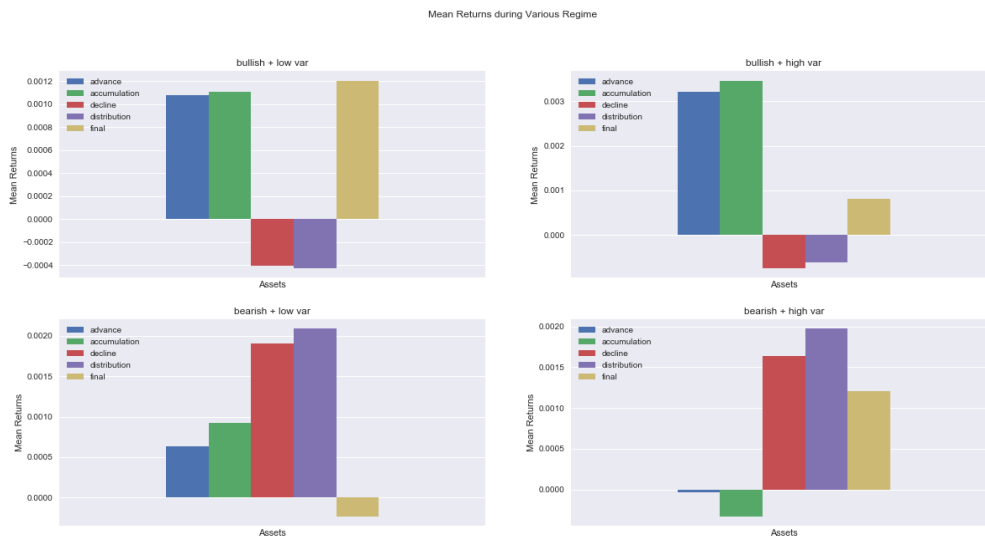

*Figure 26 Alpha performance in different regimes*

# 8. CONCLUSION

We firmly establish the importance of regime shifts models in building a systematic strategy that is tailored for all seasons. The regime shift model builds on the simple maths of the autoregressive models for time series prediction and adds a regime change component based on Markov models and when clubbed with Moving Average Regimes gives a stable model which leads to stable clearly demarcated and intuitive regimes. With this clear picture of explicit regimes in the market, the importance of adaptation of asset choices and trading styles comes out intuitively. If the returns followed one simple regime it would be very simple to model them and extract the best returns out of them, but the market changes phases, wherefore to build a dynamically adaptive alpha that can perform well in any given time, it is important to understand the dynamics of regime changes in the markets. With this new dimension added to the understanding of market patterns a systematic trading strategy can drastically enhance its risk adjusted performance, as is demonstrated in this study. We here studied a narrow subset of trading signals and asset class dynamics to evaluate the building blocks of dynamically adaptive trading system and still found very dramatic enhancements in the trading system performance. Traders come up with enhanced alphas based on novel signals, data sources, machine learning algorithms and participation rates to beat the market consistently, adding this dimension of regime adaptability can enhance the performance of the trading systems to yet another level and add stability and robustness to any trading system performance.

# APPENDIX

## I. Data Sources

| S&P 500 Index | https://finance.yahoo.com/quote/%5EGSPC/history?p=%5EGSPC |
|---|---|
| Growth | https://us.spindices.com/indices/equity/sp-500-growth |
| Consumer Discritionary | https://us.spindices.com/indices/equity/sp-500-consumer-discretionary-sector |
| Government Securities (Long term) | https://www.treasury.gov/resource-center/data-chart-center/interest-rates/Pages/Historic-LongTerm-Rate-Data-Visualization.aspx |
| Gold | https://in.investing.com/etfs/spdr-gold-trust |
| Materials | https://us.spindices.com/indices/equity/sp-500-materials-sector |
| Consumer Staples | https://us.spindices.com/indices/equity/sp-500-consumer-staples-sector |
| Banks | http://us.spindices.com/indices/equity/sp-banks-select-industry-index |
| Health | https://us.spindices.com/indices/equity/sp-500-health-care-sector |
| Value | https://us.spindices.com/indices/equity/sp-500-value |
| Real Estate | https://us.spindices.com/indices/equity/sp-500-real-estate-index |
| MSCI World | https://in.investing.com/etfs/ishares-msci-world |
| Government Securities (Short term) | https://www.treasury.gov/resource-center/data-chart-center/interest-rates/Pages/Historic-LongTerm-Rate-Data-Visualization.aspx |
| Utilities | http://us.spindices.com/indices/equity/sp-500-utilities-sector |
| Energy | https://us.spindices.com/indices/equity/sp-500-energy-sector |
| MSCI Emerging Markets | https://in.investing.com/etfs/ishares-msci-emg-markets |
| IT | https://us.spindices.com/indices/equity/sp-500-information-technology-sector |
| Oil | https://in.investing.com/commodities/crude-oil-historical-data |

## II. Performance Stats

| CAGR % | Compounded Annual Growth Rate |
|---|---|
| Annualized Risk % | Annualized Risk of the Portfolio |
| Sharpe % | Annualized Sharpe |
| Win % | Percentage of trades that gave positive returns |
| Number of Trades | Number of trades |
| Win to Loss Ratio % | Positive Trades/Loss Making trades |
| Daily PL $ | Average daily profit or loss |
| Daily Return bps (total) | Average daily return in basis points |
| Max Consecutive Losers | Maximum number of consecutive loss making trades |
| Max Drawdown % | Maximum drawdown from the peak |
| Lake Ratio | Ratio of all drawdown to returns |
| Gain to Pain Ratio | PL from profit making trades/PL from loss making trades |